\pdfoutput=1 
\documentclass[prl, aps,amsmath,groupedaddress]{revtex4}
\usepackage{graphicx}% Include figure files
\usepackage{dcolumn}% Align table columns on decimal point
\usepackage{bm}% bold math
\usepackage[english]{babel}
\begin{document}
\title{Experimental observation of the X-shaped near field spatio-temporal correlation of ultra-broadband twin beams}
\author{O. Jedrkiewicz$^{1,2}$, A.~Gatti$^{1,2}$\footnote{email:alessandra.gatti@mi.infn.it}, E.~Brambilla$^{2}$, and P.~Di Trapani$^{2}$}

\affiliation{$^1$ CNR, Istituto di Fotonica e Nanotecnologie, Piazza L. da Vinci 32 Milano, Italy,\\
$^2$ CNISM and Dipartimento di Alta Tecnologia Universit\`a dell'Insubria, Via Valleggio 11 Como, Italy}

\begin{abstract}
In this work we present the experimental observation of the non factorable near field spatio-temporal correlation of ultra-broadband twin beams generated by parametric down conversion (PDC), in an interferometric-type experiment using sum frequency generation, where both the temporal and spatial degrees of freedom of PDC light are controlled with very high resolution. The revealed X-structure of the correlation is in accordance with the predictions of the theory.
\end{abstract}
%\centerline{Version \today   }
\maketitle
\begin{figure}[h!]
\includegraphics[width=12cm]{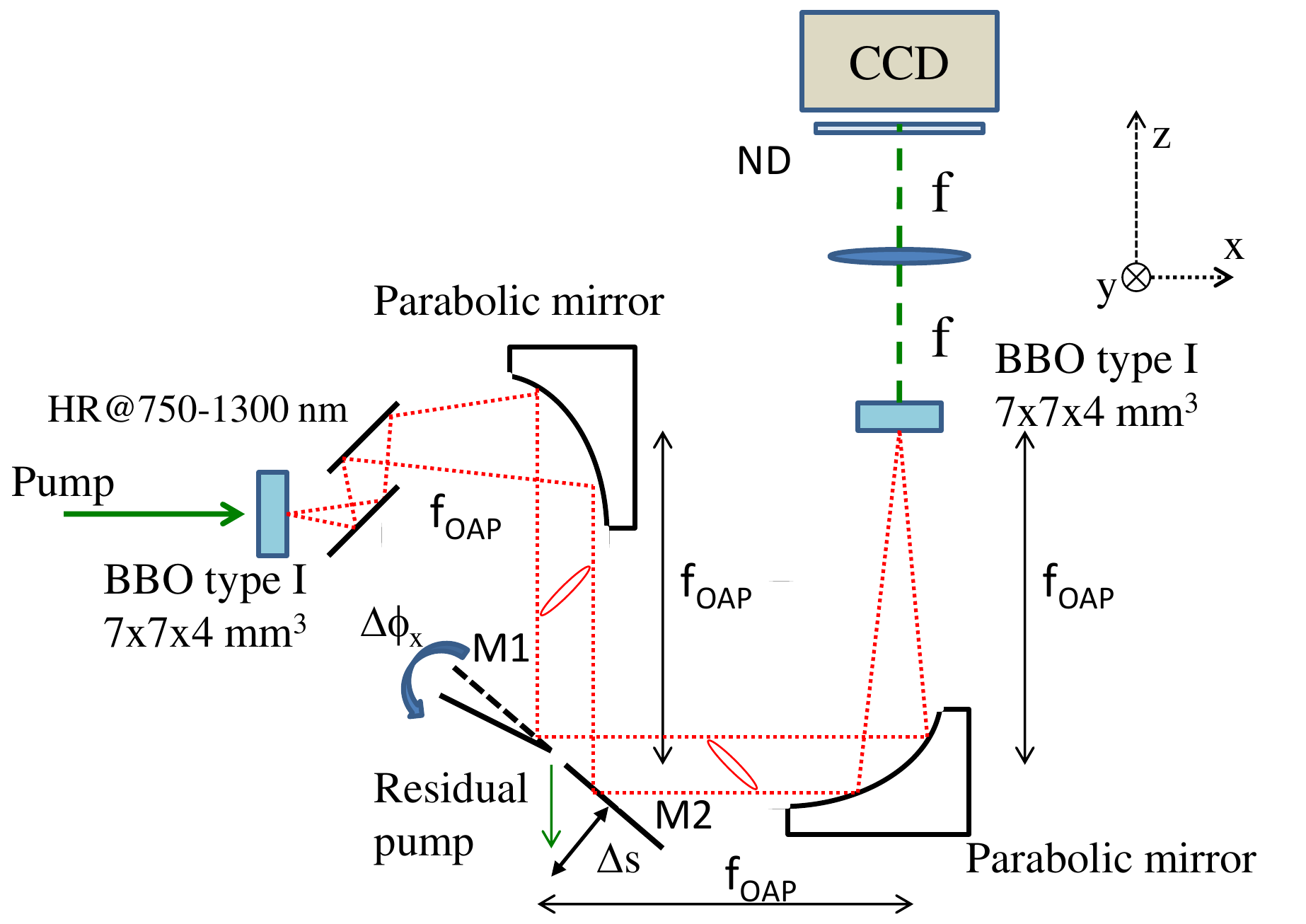}
\caption{Experimental set-up used for the investigation of the non factorable spatio-temporal correlation of the twin beams.}
\label{fig_1}
\end{figure}
\begin{figure}[h!]
\includegraphics[width=9.5cm]{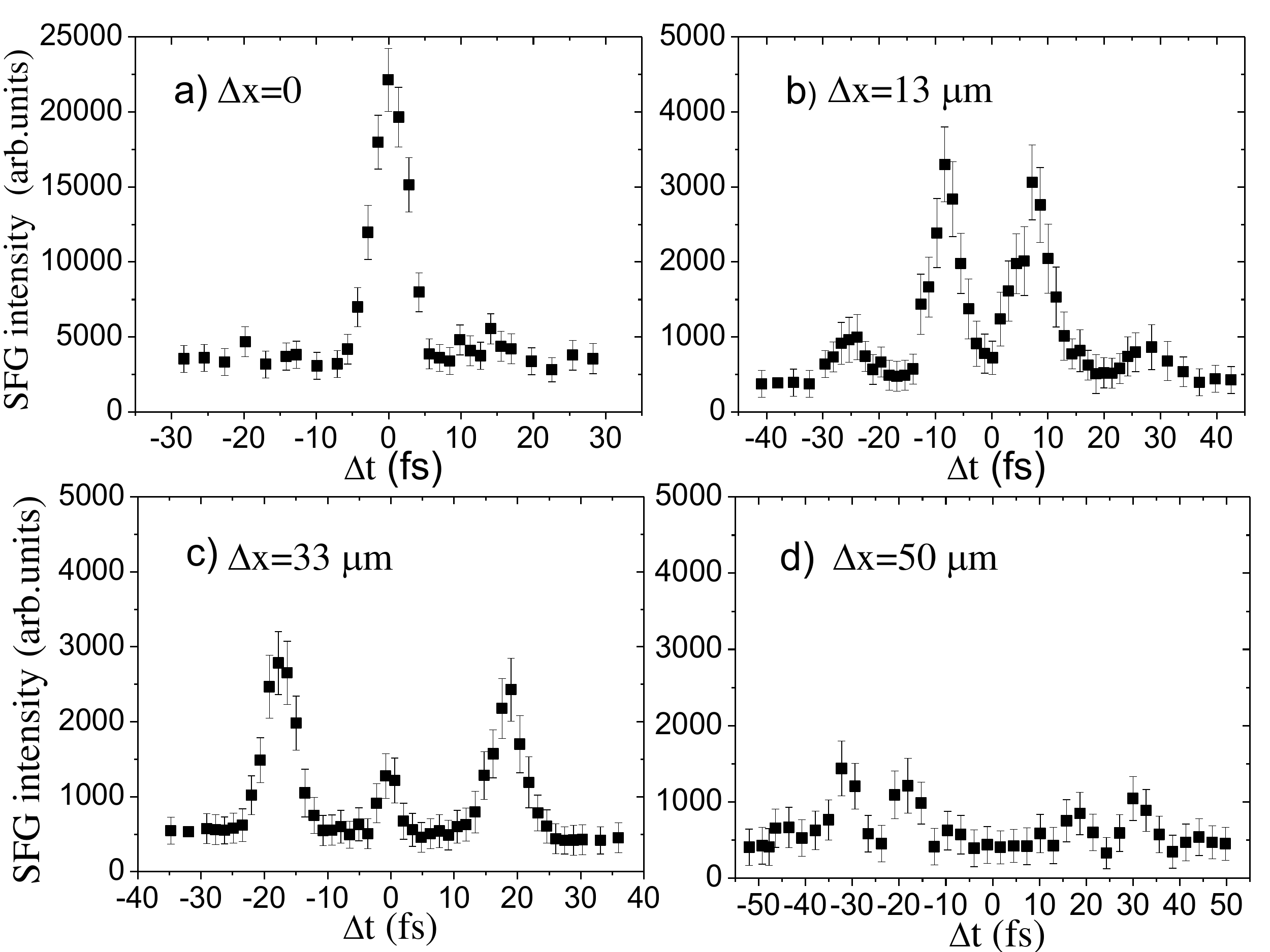}
\caption{Temporal correlation profiles of twin beams reconstructed by monitoring the SFG peak intensity as a function of a temporal delay between the beams, and plotted for four different relative transverse spatial shifts $\Delta x$ at the output plane of the PDC crystal.}
\label{fig_2}
\end{figure}
\begin{figure}[h!]
\includegraphics[width=9.5cm]{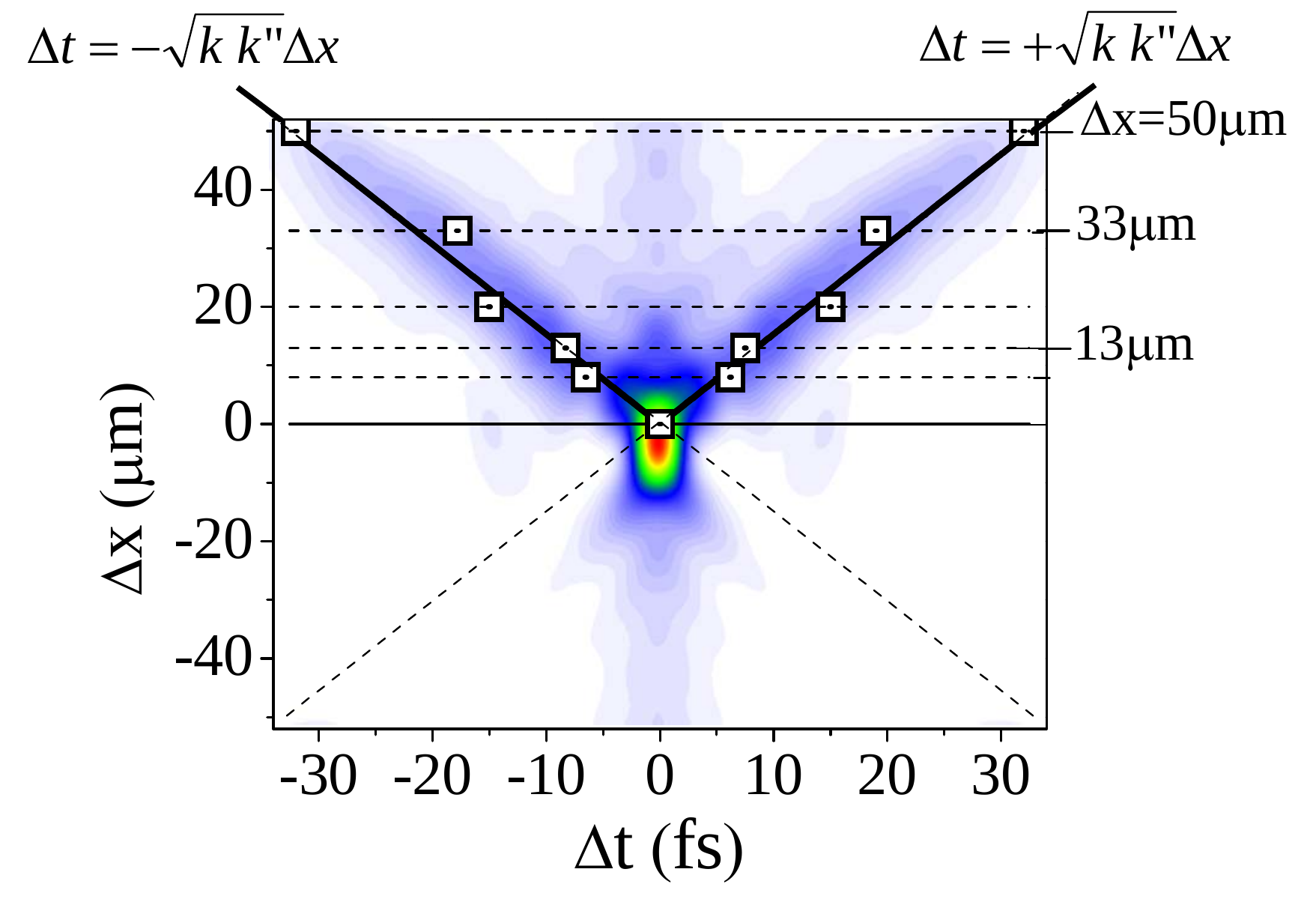}
\caption{(Color online) The density plot is the theoretical prediction for the measured SFG intensity $I_{SFG}(\Delta x,\Delta t)$, according to the model described in \cite{Brambilla12}, and with parameters close the experiment. The white squares are experimental data points showing the positions of the peaks of the measured correlation (set of data include those shown in Fig.2.)}
\label{fig_3}
\end{figure}

Spatial and temporal aspects of non linear optical phenomena are often considered separately, but when dealing with broadband phenomena where the presence of angular dispersion plays an important role, a correct description of the physical processes requires accounting for space-time coupling. This has become clear in many fields of nonlinear optics, where recent studies have revealed a rich variety of phenomena that cannot be predicted within the traditional separation between spatial and temporal degrees of freedom. This concerns for instance the study and generation of spatiotemporal solitons \cite{Wise02}, of light bullets in arrays of waveguides \cite{Minardi10}, or the generation of X-waves \cite{Ditrapani03,Kolesik04,Faccio06}.
\newline\indent
The PDC process in a nonlinear crystal, now used for several quantum information and communication schemes, has been studied for more than forty years in different contexts but usually considering separately space and time. In the first experimental work on PDC, Burnham and Weinberg \cite{Burnham70} showed that the down-converted signal and idler photons appear simultaneously within the resolving time of the detectors and the associated electronics. The \emph{temporal} correlations were studied and utilized in the pioneering series of work of Mandel and coworkers (see e.g. \cite{Mandel}), giving rise to the use of the term "twin photons", since the temporal correlation indicated that pairs of photons are born simultaneously. In general, in the past few decades, much attention has been devoted to studying the temporal two-photon interference effects involving the signal and idler photons, typically manipulated in the far-field of the nonlinear crystal (see e.g. \cite{Pittman96,Brendel91}). On the other hand, from the \emph{spatial} point of view, transverse spatial effects observed in PDC photon pairs  have provided interesting testing ground for studies of non classical states of light, quantum imaging and quantum information \cite{Gatti08, Walborn10}.  These include the demonstration of the quantum nature of the purely spatial correlation of the photon pairs, both in the far-field and near-field planes from the source \cite{Howell04,Brambilla04,Jedrkiewicz04,Moreau12,Padgett12} or of the quantum correlation in
the angular momentum and angular position of photons (see e.g. \cite{Leach10}).
\newline
\indent
The question "Do the signal and idler photons exit at the same time and at the same position from the output face of the nonlinear crystal?" would typically receive the answer "yes" or "yes, within some spatial and temporal uncertainty"
also as a consequence of the above described separable spatio-temporal picture.  In fact, in quadratic processes such as PDC, the space-time coupling, which follows directly from phase-matching (PM) where angles and frequencies are coupled through angular dispersion, plays a key role. For instance the hyperbolic geometry of the gain curve of the process in the transverse wave vector-frequency domain implies a peculiar non factorable geometry of the spatio-temporal coherence properties of the generated radiation, as shown in \cite{Jedrkiewicz06,Jedrkiewicz07} from far field spectral measurements. Recently, theoretical predictions have also highlighted in collinear PM conditions, the X-shaped spatio-temporal biphoton correlation (so called X entanglement) at the crystal exit face, presenting in low as well as high gain regime a central narrow peak (both in time and in space) and skewed arms \cite{Gatti09,Caspani10,Brambilla10}.
\newline
\indent
It seems therefore that it is only when we use a model where space and time are simultaneously taken into account that we can correctly answer the raised question. More precisely the answer should be "yes at the crystal output face the two photons can be found at the same place and at the same time within a narrow uncertainty; but one can also find, even if with lower probability, the "twin photons" relatively displaced in space by a certain amount, provided they arrive with a proportional time delay".
Nevertheless a direct verification of this X-shaped near field correlation function has to date never been made, as experiments on signal/idler or twin beam correlations have always been performed, as mentioned before, either in the spatial or in the temporal domain.
\newline
\indent
In this paper we present -to the best of our knowledge- the first \emph{experimental observation} of the X-type non factorable correlation, in space and time, of broadband twin photons.
The challenge of the experiment, performed in the high gain regime, is the ability to independently control the spatial and the temporal degrees of freedom of PDC light with a resolution in the micrometer and femtosecond ranges, respectively. The goal is achieved by means of a nonlinear interferometric-type scheme based on the inverse process of PDC that is sum frequency generation (SFG).
The SFG process has been used in classical and quantum optics as an ultrafast temporal correlator, e.g. to probe the temporal correlation of twin photons or twin beams \cite{Dayan2005,O'Donnel09,Sensarn10,Jedrkiewicz12}. In this work we investigate the correlation both \emph{in space and in time} by monitoring, for different time delays and spatial separations of the signal and idler beams, the peak intensity of the coherent pump reconstructed via sum frequency mixing of the PDC radiation.
\newline
\indent
The experimental layout used here is shown in Fig.\ref{fig_1}. The possibility of resolving the spatio-temporal shape of the correlation relies on an achromatic imaging, which avoids dispersive optical elements that would drastically deteriorate the shape of the measured correlation \cite{Brambilla12}.
%%%%%%%%%%%%%%%%%%%%%%%%%%%%%%%%%%%%%%%%%%%%%%%%%%%%%%%%%%%%%%%%%%%%%%%%%%%%%%%%%%%%%%%%%%%%%%%%%%%%%%%%%
\newline
 The pump pulse at 527.5 nm, obtained from the second harmonic of a 1 ps, 1055 nm, 10 Hz repetition rate Nd:Glass laser (Twinkle, Light Conversion Ltd.), is collimated down to about 0.8 mm (FWHM) at the entrance of a type I 4 mm Beta Barium Borate (BBO) crystal for PDC in the collinear configuration. Just after the crystal, two custom made high reflectivity dielectric mirrors (Layertec) with reflectivity $R>99\%$ in the 730-1300 nm range and $R\approx0.3\%$ at $\lambda=527.5nm$, are used to reflect the PDC radiation. The achromatic imaging from the output plane of the PDC crystal is performed by means of two identical 90 degree off-axis parabolic gold mirrors (OAP) collecting a huge portion of the spatio-temporal spectrum of the PDC radiation ($\geq 600 nm$), as illustrated in \cite{Jedrkiewicz12}. The frequency mixing of the PDC radiation is performed in a second BBO crystal identical to the first. This SFG crystal is placed in the focal plane of the second OAP mirror and is mounted on a micrometer translation stage permitting to finely adjust its position relatively to the imaging plane of the system. Both crystals are also mounted on a rotation stage in order to adjust their orientation with the aim of working at exactly the same PM conditions.
\newline\indent
A detailed theoretical analysis of a scheme analogous to this one is presented in Ref. \cite{Brambilla12}, which shows  how the full spatio-temporal structure of the PDC correlation can be disclosed by using the SFG process, by inserting in a controlled way a temporal delay $\Delta t$ and a transverse spatial displacement $\Delta \vec{x}$ between twin beams. In the experiment this  is realized  by means of the two adjacent gold mirrors $M_1$ and $M_2$ (see Fig.\ref{fig_1}) placed in the far-field plane of the PDC crystal. They act separately on the twin  components of the light, because each photon (signal) has its twin (idler) on the opposite side of the far-field  due to momentum conservation in the elementary PDC process.  While mirror $M_2$ can be translated by an amount $\Delta s$ to produce a relative delay $\Delta t  = \sqrt{2} \Delta s /c$ between twin beams, the mirror $M_1$ can be rotated by an angle $\Delta\phi$ in order to generate at the SFG crystal input face a transverse displacement of one beam with respect to the other by an amount equal to $2f_{OAP}$ $\Delta\phi$. A 2mm wide gap between the two mirrors allows to eliminate the residual input pump. Since the correlation is strongly localized both in space and in time (in the micrometer and femtosecond range respectively), the temporal and spatial relative displacements of the twin beams must be scanned with micrometric precision. This is done thanks to high resolution translation and rotation stage piezo-controllers (Mercury, PI).
\newline\indent
For what concerns the diagnostics, the radiation emitted by the SFG crystal is observed in the focal plane of a f=20 cm focal length lens and the light intensity is monitored by means of a 16 bit scientific CCD camera (Roper Scientific) with 80 $\%$ detection efficiency at 527.5 nm.
In this plane we observe  a narrow  light peak originating from coherent up-conversion processes, and reproducing the far-field profile of the original pump \cite{Jedrkiewicz11}, lying over a widely spread incoherent speckle field constituting the background.
\newline\indent
The intensity of the coherent peak,  monitored as a function of $\Delta \vec{x}$ and $\Delta t$ is able to give a precise information about the spatio-temporal correlation of twin beams at the exit face of the PDC crystal, namely about the modulus of the biphoton correlation $\Psi(\Delta\vec{x},\Delta t): =
<A(\vec{x}, t) A(\vec{x} + \Delta \vec{x}, t+\Delta t>$, where $A$ is the PDC field operator at the crystal output \cite{Brambilla12}. This quantity represents the probability amplitude that a photon pair is found at the crystal exit face separated by $\Delta \vec{x}$ and delayed by an amount $\Delta t$. In conditions close to collinear PM and for a broad enough pump beam, such biphoton correlation assumes an X-shape in any plane containing the temporal delay $\Delta t$ and one spatial coordinate $\Delta x$, expressing a strong relationship between the temporal delays and the spatial separations of twin photons at the crystal output face:
\begin{equation}
\Delta t = \pm \sqrt{k \, k''} \Delta x
\label{dtdx}
\end{equation}
where $k'' = \left. \frac{d^2 k}{d\Omega^2} \right|_{\Omega=0}$ is the group velocity dispersion (GVD) coefficient.
\newline\indent
The experimental results, obtained thanks to a careful control of the alignment criticity, are presented in Figs.\ref{fig_2} and \ref{fig_3}. We mention that for technical reasons the mirror $M_1$ has been mounted in such a way that the rotation occurs around the "gap" axis between $M_1$ and $M_2$ (the y axis, orthogonal to the plane of Fig.\ref{fig_1}). In this configuration,  the displacement produced between the two (semi-circular) beams lies along the x-direction in the plane of Fig.\ref{fig_1}. This choice affects the measured reconstructed correlation
%(in our case denoted by $|\Psi_{exp}(\Delta x,0,\Delta t)|^{2}$),
whose shape as discussed in \cite{Brambilla12} changes from an X to a V-shape with vanishing correlation for negative displacements $\Delta x<0$, and enhanced visibility of its skewed tails for $\Delta x>0$.
In fig. \ref{fig_2}, each plot, representing a temporal correlation profile, is associated with a given transverse spatial separation of the twin beams at the SFG crystal input face (i.e. at the imaging plane of the output face of the PDC crystal). Here we present the twin beam temporal correlation profiles obtained for four positive values of $\Delta x$ by monitoring the SFG peak intensity as a function of the delay $\Delta t$ introduced between the two beams. Each data point corresponds to the coherent peak intensity averaged over 15 images each of them recorded over 2s (20 laser shots). The single temporal correlation peak, observed  for $\Delta x=0$, splits  for $\Delta x > 0$ into two peaks symmetrical with respect to the $\Delta t = 0$ position, and whose temporal separation increases for increasing values of $\Delta x$. The correlation time that characterizes the width of the two peaks is close to that found for the single central peak at $\Delta x$ = 0, about 6fs at FWHM.
Note that the small central peak visible in Fig.\ref{fig_2}c, also present in the theoretical density plot displayed in Fig.\ref{fig_3}, is a consequence of the choice of the rotation axis of mirror $M_2$, and would not be present for displacements along the orthogonal direction $\Delta y$. In Fig.\ref{fig_3} the positions of the correlation peaks obtained from an entire set of measurements, are reported in the ($\Delta x$,$\Delta t$)-plane,
showing how they lie along the asymptotes of the X-structure (black bold lines) given by  $\Delta t = \pm \sqrt{k \, k''} \Delta x$ (see Eq.\eqref{dtdx}).

%%%%%%%%%%%%%%%%%%%%%%%%%%%EXP RESULT Dx> 0 %%%%%%%%%%%%%
%%%%%%%%%%%%%%%%%%%%%%%%%%%%%%%%%%%%%

%%%%%%%%%%%%%%%%%%%FIGURA CON LA X %%%%%%%%%%%%%%%%%%%%%%%%%%%%%%%%%%%%%%%%%%%%%%%%%%%%%%
%%%%%%%%%%%%%%%%%%%%%%%%%%%%%%%%%%%%%%%%%%%%%%%%%%%%%%

This relation has its very origin in the phase matching mechanism that imposes a balance between the GVD, responsible for the relative delay acquired by the twin photons during propagation, and the diffraction, which causes their relative transverse separation \cite{Gatti12}. In order to elucidate this point, let us consider closely the mechanism of photon pairs generation. Energy-momentum conservation imposes that, in the limit of a plane-wave and monochromatic pump, twin photons are generated with opposite transverse wave-vectors $\pm \vec{q}$, and  with symmetric frequency offsets $\pm \Omega$ with respect to the central frequency $\omega_0/2$. On the other hand, efficient down-conversion occurs only for those photon pairs satisfying longitudinal PM, i.e. for which $ k_z (\vec{q}, \Omega) + k_z (-\vec{q}, -\Omega) -k_0 \approx 0$, where $k_z$ is the longitudinal component of their wave-vector and $k_0$ the pump wave-number. By making a Taylor expansion up to second order, the PM in collinear conditions can be recasted as \cite{Caspani10}
\begin{equation}
q^2 = k\, k'' \Omega^2 \ ,
\label{PM2}
\end{equation}
expressing the well known features of the type I fluorescence cones, where photon pairs propagating at larger angles with respect to the pump have larger frequency offsets.
Let us now assume that a pair of photons is generated  at any point in the crystal at a longitudinal distance z from its output.
During propagation the two members of the pair can be delayed one with respect to the other because of GVD. Interpreting the two photons as two wave-packets  centered around $\pm \Omega$,  with group velocities $v_g(\Omega) = dk (\Omega) /d\Omega$, the relative delay between twin photons  at the crystal output face is  $\Delta t (z, \vec{q}, \Omega) \approx  z/v_g (\Omega) - z/v_g(-\Omega) \approx 2 k'' \Omega$, where we assumed small angles and frequency offsets (same limit of validity as \eqref{PM2}).
Because of diffraction, twin photons separate in space: assuming they are e.g. emitted in the (x,z) plane,
and using again the analogy with classical wave packets,
their relative separation at the crystal exit is $ \Delta x (z, \vec{q}, \Omega)=
z \tan{\theta(q,\Omega) } -z\tan{\theta(-q,-\Omega) } \approx 2 z q/k$ where $\theta$ is the angle between the propagation direction of the photon  and the z-axis, and we assumed small angles, i.e. $\tan{\theta(q,\Omega) } \approx q/k_z(q, \Omega) \approx q/k$.
By using now the PM relation \eqref{PM2}, we easily obtain $\Delta t (z, \vec{q}, \Omega) = \pm \sqrt{k \, k''} \Delta x (z, \vec{q}, \Omega)$, valid irrespectively of the point where the photon pair was created and of the mode $\vec{q},\Omega$, implying thus the proportionality between temporal delays and spatial separations expressed by \eqref{dtdx}. We remark that this relation has a much wider range of validity than that derived here, as clarified in \cite{Gatti12}.
\newline
\indent
In conclusion, we have experimentally demonstrated the non factorability of the spatio-temporal correlation of twin beams in a high gain PDC experiment.
These results, having their origin in the PM mechanism of the process, which imposes a balance between GVD and diffraction, suggest that signal and idler photons do not necessarily exit from the PDC crystal at the same time and in the same place, but that their emission out of the non linear medium is correlated in the space-time domain along skewed lines.
  The result is once more a clear evidence of the importance of accounting for space-time coupling in all those broadband processes where the presence of angular dispersion plays a key role. Note that the revealed X-type spatio-temporal correlation presents a geometry very similar to the classical first order coherence of the PDC field, inferred in Ref.\cite{Jedrkiewicz06} from the measured far field spatio-temporal spectrum.
  In contrast, in the present work, the X-type \emph{twin beam correlation} has been \emph{directly demonstrated} in the space-time domain, via a nonlinear interferometric type experiment based on SFG, that guarantees the photons phase-conjugation over a huge bandwidth, necessary condition for resolving the narrow temporal peaks of the correlation.
\newline
\indent
O.J. thanks J.-L. Blanchet for help in the mounting stage of the experimental set-up. The authors acknowledge support from Grant No. 221906 HIDEAS of the Fet Open Programme of the EC.


\begin{thebibliography}{}

\bibitem{Wise02} see F. Wise and P. Di Trapani, 
%"The Hunt for Light Bullets - Spatiotemporal Solitons", 
Optics and Photonics News, February 2002, and references therein.
\bibitem{Minardi10} S. Minardi et al., Phys. Rev. Lett, \textbf{105}, 263901 (2010).
\bibitem{Ditrapani03} P. Di Trapani, G. Valiulis, A. Piskarskas, O. Jedrkiewicz, J. Trull, C. Conti, and S. Trillo,  Phys. Rev. Lett. \textbf{91}, 93904 (2003).
\bibitem{Kolesik04} M. Kolesik, E. M. Wright, and J. V. Moloney, Phys. Rev. Lett. \textbf{92}, 253901 (2004).
\bibitem{Faccio06} A. Couairon, E. Gaižauskas, D. Faccio, A. Dubietis, and P. Di Trapani, Phys. Rev. E \textbf{73}, 16608 (2006).
\bibitem{Burnham70} D. C. Burnham and D. L.Weinberg, Phys. Rev. Lett. \textbf{25}, 84 (1970).
\bibitem{Mandel}  C.K. Hong, L. Mandel, Phys. Rev. Lett. \textbf{56}, 58 (1986); R. Ghosh, L. Mandel, Phys. Rev. Lett. \textbf{59}, 1903 (1987), and references therein.
\bibitem{Pittman96} T. B. Pittman, D. V. Strekalov, A. Migdall, M. H. Rubin, A. V. Sergienko, and Y. H. Shih, Phys. Rev. Lett. \textbf{77}, 1917 (1996).
\bibitem{Brendel91} J. Brendel, E. Mohler, and W. Martienssen, Phys. Rev. Lett. \textbf{66}, 1142 (1991).
\bibitem{Gatti08} Gatti, E. Brambilla, and L. A. Lugiato, Progress in Optics \textbf{51}, 251 (2008), and references therein
\bibitem{Walborn10} S.P. Walborn, C.H. Monken, S. Pádua, P.H. Souto Ribeiro, Physics Reports \textbf{495}, 87 (2010), and references therein.
\bibitem{Howell04} J. C. Howell, R. S. Bennink, S.J. Bentley and R.W. Boyd, Phys. Rev. Lett 92, 210403(2004)
\bibitem{Brambilla04} E. Brambilla, A. Gatti, M. Bache, and L. Lugiato, Phys. Rev. A \textbf{69}, 023802 (2004)
\bibitem{Jedrkiewicz04} O. Jedrkiewicz, Y.-K Jiang, E. Brambilla, A. Gatti, M. Bache, L. A. Lugiato, and P. Di Trapani, Phys. Rev. Lett. \textbf{93}, 243601 (2004).
\bibitem{Moreau12} P.-A. Moreau, J. Mougin-Sisini, F. Devaux, and E. Lantz, Phys. Rev. A \textbf{86}, 010101(R) (2012).
\bibitem{Padgett12} M. P. Edgar et al., arXiv:1204.1293v1 (2012).
\bibitem{Leach10} J. Leach, B. Jack, J. Romero, A. K. Jha, A. M. Yao, S. Franke-Arnold, D. Ireland, R. W. Boyd, S. M. Barnett, M. J. Padgett, Science {\bf 329}, 662 (2010).
\bibitem{Jedrkiewicz06} O. Jedrkiewicz, A. Picozzi, M. Clerici, D. Faccio, and P. Di Trapani, Phys. Rev. Lett. \textbf{97}, 243903 (2006).
\bibitem{Jedrkiewicz07} O. Jedrkiewicz, M. Clerici, A. Picozzi, D. Faccio, and P. Di Trapani, Phys. Rev. A \textbf{76}, 033823 (2007).
\bibitem{Gatti09} A. Gatti, E. Brambilla, L. Caspani, O. Jedrkiewicz, and L. A. Lugiato, Phys. Rev. Lett. \textbf{102}, 223601 (2009).
\bibitem{Caspani10} L. Caspani, E. Brambilla and A. Gatti, Phys. Rev. A\textbf{ 81}, 033808 (2010).
\bibitem{Brambilla10} E. Brambilla, L. Caspani, L.A. Lugiato, and A. Gatti, Phys. Rev. A \textbf{82}, 013835 (2010).
\bibitem{Dayan2005}	 A. Pe'er, B. Dayan, A. Friesem, and Y. Silberberg  Phys. Rev. Lett {\bf 94}, 073601  (2005).
\bibitem{O'Donnel09} K. A. O' Donnel and A. B. U'Ren., Phys. Rev. Lett {\bf 103}, 123602 (2009).
\bibitem{Sensarn10} S. Sensarn, G.Y. Yin, and S. E. Harris. Phys. Rev. Lett. \textbf{104}, 253602 (2010).
\bibitem{Jedrkiewicz12} O. Jedrkiewicz, J.-L. Blanchet, E. Brambilla, P. Di Trapani, and A. Gatti, Phys. Rev. Lett. \textbf{108}, 253904 (2012).
\bibitem{Brambilla12} E. Brambilla, O. Jedrkiewicz, L. A. Lugiato, and A. Gatti, Phys. Rev. A \textbf{85}, 063834 (2012).
\bibitem{Jedrkiewicz11} O. Jedrkiewicz, J.-L. Blanchet, A. Gatti, E. Brambilla and P. Di Trapani, Opt. Exp. \textbf{19}, 12903 (2011).
\bibitem{Gatti12} A.Gatti, L. Caspani, E. Brambilla,  An intuitive picture of the X-correlation of twin photons , preprint.
\end{thebibliography}
\end{document}